# Tailoring Magnetic Exchange Interactions in Ferromagnet-Intercalated MnBi$_2$Te$_4$ Superlattices

Peng Chen[1,2,3], Qi Yao[4,5*], Qiang Sun[6,7], Alexander J. Grutter[8], P. Quarterman[8], Purnima P. Balakrishnan[8], Christy J. Kinane[9], Andrew J. Caruana[9], Sean Langridge[9], Baoshan Cui[10,11], Lun Li[1,2,3], Yuchen Ji[5], Yong Zhang[1,2,3], Zhongkai Liu[4,5], Jin Zou[6,7], Guoqiang Yu[10,11], Yumeng Yang[1*], and Xufeng Kou[1,4*]

[1]School of Information Science and Technology, ShanghaiTech University, Shanghai, 201210, China

[2]Shanghai Institute of Microsystem and Information Technology, Chinese Academy of Sciences, Shanghai 200050, China

[3]University of Chinese Academy of Science, Beijing 101408, China

[4]ShanghaiTech Laboratory for Topological Physics, ShanghaiTech University, Shanghai 201210, China

[5]School of Physical Science and Technology, ShanghaiTech University, Shanghai, 201210, China

[6]School of Mechanical and Mining Engineering, The University of Queensland, St Lucia, QLD 4072, Australia

[7]Centre for Microscopy and Microanalysis, The University of Queensland, St Lucia, QLD 4072, Australia

[8]NIST Center for Neutron Research, National Institute of Standards and Technology, Gaithersburg, Maryland 20899-6102, USA

[9]ISIS-Neutron and Muon Source, Rutherford Appleton Laboratory, Didcot, Oxon OX11 0QX, United Kingdom

[10]Songshan Lake Materials Laboratory Dongguan, Guangdong 523808, China

[11]Beijing National Laboratory for Condensed Matter Physics Institute of Physics Chinese Academy of Sciences Beijing 100190, China

* Corresponding author. Email: yaoqi@shanghaitech.edu.cn; yangym1@shanghaitech.edu.cn; kouxf@shanghaitech.edu.cn;

**The intrinsic magnetic topological insulator MnBi$_2$Te$_4$ (MBT) has provided a platform for the successful realization of exotic quantum phenomena. To broaden the horizons of MBT-based material systems, we intercalate ferromagnetic MnTe layers to construct the [(MBT)(MnTe)$_m$]$_N$ superlattices by molecular beam epitaxy. The effective incorporation of ferromagnetic spacers mediates the anti-ferromagnetic interlayer coupling among the MBT layers through the exchange spring effect at the MBT/MnTe hetero-interfaces. Moreover, the precise control of the MnTe thickness enables the modulation of relative strengths among the constituent magnetic orders, leading to tunable magnetoelectric responses, while the superlattice periodicity serves as an additional tuning parameter to tailor the spin configurations of the synthesized multi-layers. Our results demonstrate the advantages of superlattice engineering for optimizing the magnetic interactions in MBT-family systems, and the ferromagnet-intercalated strategy opens up new avenues in magnetic topological insulator structural design and spintronic applications.**

In magnetic topological insulators (MTIs), the coexistence of broken time-reversal symmetry (TRS) and robust topologically protected surface states gives rise to a variety of emergent TRS-breaking physics[1-5]. One successful method to obtain MTIs is the incorporation of transition-metal elements (Cr, V, Mn) into the host TI matrix to establish the robust ferromagnetism[2,6,7]. However, the random distribution of magnetic atoms inevitably leads to a spatial fluctuation of the magnetic exchange gap, which restricts MTI-related phenomena at deep cryogenic temperatures. One approach to address such challenge is the design of TI-based magnetic heterostructures in which uniform magnetic order can be induced by an interfacial proximity effect[8]. In this way, the separation of the topology and the magnetism into different layers allows us to control their contributions independently. However, due to critical lattice-matching requirements, the selection of TI-compatible magnetic films is limited and only a few MTI heterostructures have been fabricated[9-11].

Alternatively, MnBi$_2$Te$_4$ (MBT), an intrinsic MTI material, has aroused extensive attention in recent years[12-14]. In contrast to magnetic doping, the Mn atoms in MBT strictly form a two-dimensional lattice plane within the stacked Te-Bi-Te-Mn-Te-Bi-Te septuple layers (SL). This structure not only preserves the large spin-orbit coupling required for band inversion, but also provides a homogeneous magnetic ground-state with intra-layer ferromagnetism (FM) and inter-layer A-type anti-ferromagnetism (AFM)[15]. As a result, dissipationless chiral edge conduction has been observed in thin MBT flakes above 1 K, although it requires the presence of a large external magnetic field to fully magnetize the samples[12,16,17].

In this regard, new breakthroughs may be found through deliberate modifications of the MBT framework, from which spontaneous magnetization can be achieved at zero magnetic field[12]. For instance, recent attempts by inserting the Bi$_2$Te$_3$ spacer in (MBT)(Bi$_2$Te$_3$)$_n$ ($n$ = 1-6) series crystals have been made, where the inter-layer AFM-type Anderson super-exchange coupling strength is reduced significantly with increasing $n$[18-22]. Furthermore, instead of this non-magnetic intercalation strategy, which may cause unintended magnetic fluctuations[23], Mn$_2$Bi$_2$Te$_5$ has been proposed as a superior candidate given that the additional Mn layer in the Mn-Te double spacer improves the magnetic stability of the original MBT SL. Unfortunately, during the Mn$_2$Bi$_2$Te$_5$ single crystal growth, the Mn atoms tend to dope into MBT rather than forming the stable Mn-Te-Mn structure[15], hence the experimental realization of FM-intercalated MBT proposals remains elusive[24].

In this work, we utilize the molecular beam epitaxy (MBE) technique to integrate ferromagnetic MnTe and MBT into [(MBT)(MnTe)$_m$]$_N$ superlattices, where each of the $N$ periods is composed of one MBT septuple layer and $m$ MnTe unit cells. The insertion of the MnTe spacer with perpendicular FM order excels in both tuning the AFM-to-FM strength and stabilizing the interlayer exchange coupling between individual MBT layers. Furthermore, through deliberate superlattice periodicity control, we also manipulate the spin texture of the entire superlattice structure with configurable magneto-resistance (MR) responses. The generic design rules of FM/MBT hybrid systems may shed light on the exploration of MTI-based physics and multi-functional device applications.

Experimentally, we first grew single-crystalline MBT films by depositing individual MnTe and $Bi_2Te_3$ layers on $Al_2O_3$(0001) substrates in an alternating sequence, followed by a moderate post-annealing procedure to form the MBT SL structure. The absence of dangling bonds on the $Al_2O_3$ surface promotes the epitaxial growth of MBT, and the sharp streaky pattern of the *in-situ* reflection high-energy electron diffraction (RHEED) reveals the two-dimensional growth mode of the as-grown MBT films (Fig. 1a). The cross-sectional high-resolution scanning transmission electron microscopy (HR-STEM) image of the MBT sample (Fig. 1b) clearly unveils the highly ordered atomic distribution of the desired Te-Bi-Te-Mn-Te-Bi-Te sequence. In addition, the quantitative analysis of the electron energy loss spectrum (EELS) in Fig. 1c confirms an elemental composition ratio of Mn:Bi:Te = 1:2:4 in the single-crystalline film, and the X-ray diffraction (XRD) pattern (Supplementary Fig. 1) exhibits a series of (00$n$) peaks without observable secondary phases, both of which agree well with the ideal stoichiometric MBT values. Through the polarized neutron reflectometry (PNR) measurements of the Te-capped MBT film, the nuclear scattering length density (SLD) profile is consistent with bulk-like MBT and the magnetic SLD diagram shows a uniform magnetization distribution across the film thickness, with a magnitude in accordance with that expected from MBT in an in-plane magnetic field, supporting high-quality growth of the designed phase (Supplementary Fig. 2-1).

Subsequently, we performed four-point magneto-transport experiments on the 5 SL MBT sample to explore its magnetic/electrical properties. Consistent with studies of exfoliated MBT flakes[12,16,25], the magneto-resistance displays a hump-like line-shape at $T$ = 1.5 K, as highlighted in Fig. 1d. In particular, the initial anti-parallel alignment of spins in adjacent MBT layers leads to the high-MR state in the low-field region, and the abrupt increase in magnitude to a MR peak near a field of ± 3.3 T (*i.e.,* defined as the transition field $H_t$) corresponds to the characteristic spin-flop feature for an *A*-type AFM system[12,16,25]. Upon further increasing the applied magnetic field, all magnetic moments become polarized along the *z*-direction at the saturation field of $H_s$ = 7 T, and the MBT sample falls into a low-MR state. This behavior is similar to the giant magnetoresistance (GMR) effect found in a spin-valve system (the detailed GMR model is discussed in Supplementary Fig. 3). Besides, the corresponding anomalous Hall resistance ($R_{xy}$) data (Fig. 1e) reaches saturation above $H_s$ with $R_s$ = 4 kΩ, a relatively larger value compared with other reported MBE-grown MBT films[15,26,27]. It is noted that the hybrid-AHE-like $R_{xy}$ slope at low magnetic fields (± 3.3 T) may be caused by native anti-site defects and/or random stacking order in MBT, which in turn modify the local electronic structures and magnetic moments[22,24,26]. From the aforementioned structural and electrical characterizations, we have validated the high quality of our MBT layer as a reliable building block for the following study.

Based on our previous experience with the FM-type MnTe film growth[28], we have chosen it as the intercalated layer to create the [(MBT)(MnTe)$_m$]$_N$ system. As illustrated in Fig. 1f, the similar 2D stoichiometric structures with a small lattice mismatch between MnTe and MBT guarantees the epitaxial growth of the superlattice structure in which the strong ferromagnetism (*i.e.,* Curie temperature of $T_C$ ~ 150 K) of MnTe can add robust FM interactions to the hybrid system, and the MnTe layer number *m* (*i.e.*, the spacing between adjacent MBT layers) can be, in principle, utilized to tune the inter-layer AFM coupling. It is noteworthy that the maximum thickness of the MnTe layer in this study is limited below 1 nm (*m* = 1.75), which is sufficiently thin to ensure effective couplings among MBT layers in the superlattice[23,29].

Given the known tendency of Mn to diffuse extensively in MBT-related systems, we further performed neutron reflectometry experiments to probe the superlattice depth and magnetic profiles. Firstly, an unpolarized measurement of the nominal *m* = 1.75 sample with *N* = 10 SL repeats yields an experimental thickness of $m = 1.72$, well within the experimental uncertainty of the expected value (Supplementary Fig. 2-2). Together with the appearance of the anticipated superlattice first order Bragg reflection peak at a very high reciprocal lattice vector of $Q \approx 2.7$ nm$^{-1}$, such supplementary information certifies the realization of large-area, highly periodic [(MBT)(MnTe)$_{1.75}$]$_{10}$ superlattice by MBE (Supplementary Fig. 2-2a). Next, an additional PNR measurement was conducted at $T = 6$ K under the applied in-plane magnetic field of 3 T. Figures 2a-b summarize the best fitting results of the non-spin-flip reflectivity and spin asymmetry signals (*i.e.,* defined as the difference between the spin-dependent reflectivities normalized by their sum) in reference to the desired [(MBT)(MnTe)$_{1.75}$]$_{10}$ superlattice structure. While this measurement did not reach the first order Bragg reflection due to the lower intensity of the polarized neutron beam, the oscillatory SLD depth profile in Fig. 2c resides within the reasonable range between the bulk MBT and MnTe values, and the integrated moment detected by PNR manifests a net contribution to the magnetization from the incorporated MnTe spacers. Assuming that the MBT moment itself does not increase progressively with the applied field, our PNR results thus indicate the resulting canted spin configuration with a non-zero in-plane magnetization of 20 emu/cc may originate from the interplay between the MBT and MnTe layers through interfacial and interlayer exchange interactions.

In order to investigate the influence of MnTe intercalation, we have conducted the temperature-dependent MR measurements on a set of [(MBT)(MnTe)$_m$]$_5$ samples with $m$ varying from 0 to 1.75 in Fig. 3. When the samples are cooled to $T = 1.5$ K (below the Néel temperature of MBT $T_N = 25$ K), the MR signals exhibit a strong $m$-dependence. As disclosed in Fig. 3a, the MR peak transition field $H_t$ monotonically shrinks with the increase of the MnTe thickness (*i.e.,* the change of $H_t$ as compared to the pure MnTe counterpart ($\Delta H_t$) successively reduces from 2.6 T ($m = 0.25$) to 0.5 T ($m = 1.75$) at $T = 1.5$ K, as marked by the red circles in Fig. 3c), signifying the effective modulation of the spin-flop effect through the MnTe intercalation. Meanwhile, the dramatic suppression of the MR amplitude from $m = 0$ (12.5%) to $m = 0.5$ (2.65%) implies the reduction of the AFM-related GMR-like behavior. By further increasing $m$, the FM contribution of the MnTe layer becomes more pronounced, therefore triggering the appearance of a double-split butterfly MR response from its original curve. Eventually, the overall MR profiles of the MnTe-dominated superlattice system ($m \geq 0.75$) resembles that of the pure MnTe control sample except for a larger transition field ($H_t$). In fact, the $m$-dependent $H_t$ identified in Fig. 3a is reminiscent of the exchange-spring magnet behavior observed in conventional magnetic multilayers[30]. Depending on the relative strengths of the anisotropy and exchange coupling energy between constituent magnetic components, the ground-state spin-texture of the system and its correlated magneto-transport responses can be tuned[31]. In our MnTe-intercalated MBT superlattices, when the soft magnet MnTe couples with the magnetically hard MBT single layer, the strong anisotropy of the latter can pin the magnetic moment within the MnTe layer via the magnetic proximity coupling at the interface[30,32-35], resulting in the enlarged transition field $H_t$. On the other hand, as the base temperature is elevated above the $T_N$ of MBT, the absence of the AFM-type inter-layer Anderson super-exchange coupling would drive the pure 5 SL MBT sample into a paramagnetic state, as verified by the positive, parabolic MR curve detected at 30 K (*i.e.,* the MBT curve of Fig. 3b). Afterwards, the entire magneto-transport behavior of the [(MBT)(MnTe)$_m$]$_5$ system is only governed by the remaining MnTe-related FM order. Consequently, typical butterfly-shape MR slopes with a constant peak position are recorded, regardless of the MnTe thickness (*i.e.,* same as the pure FM-type MnTe film[28]). Furthermore, relevant temperature-dependent $H_t$ slopes in Fig. 3d display a clear MBT-to-MnTe transition trait as $m$ varies from 0.25 to 1.75, providing a guidance in evaluating the competing intra-/inter-layer magnetic couplings.

In addition to enabling the coercivity modulation, the establishment of exchange-spring effect is also found to mediate the interaction between adjacent MBT layers in our [(MBT)(MnTe)$_m$]$_N$ superlattices. Unlike the $m$-dependent $H_t$ phenomenon noted in the low-field region, the saturation field $H_s$, which characterizes the inter-layer exchange coupling strength, remains constant at 7 T (*i.e.,* the same value as the non-intercalated MBT film) and shows a negligible variation with respect to $m$ (*i.e.,* the dashed straight line in Fig. 3a). In order to understand this distinctive feature, we carried out high-field anomalous Hall resistance measurements on the same set of [(MBT)(MnTe)$_m$]$_5$ samples at low temperatures. Strikingly, the normalized Hall resistance ($R_{xy}/R_s$) curves at $T$ = 1.5 K (Fig. 4a) shows almost identical contours, indicating a universal transition process as the magnetic moments in the superlattice are gradually aligned by the applied field. In agreement with the MR results in Fig. 3, the saturation field $H_s$ values extracted from the AHE data at $T$ =1.5 K, 3 K, 5 K, and 10 K also stay independent on the MnTe spacer thickness (Fig. 4b). Similarly, the $H_s$–$T$ slopes also follow the same temperature-scaling trace up to 25 K (Fig. 4c), above which the inter-layer AFM order disappears (*i.e.,* $H_s$ = 0) and the overall AHE signal reverts to that of a single-phase FM-driven $R_{xy}$ hysteresis loop.

Quantitatively, the distinct evolutions of $H_t$ and $H_s$ versus the MnTe thickness can be well described by a modified linear chain model[36-38], in which the magnetization of each constituent MBT and MnTe layer is considered as a 'macro spin'. As schematized in Fig. 5a, the magnetic interactions in the [(MBT)(MnTe)$_m$]$_N$ superlattice system consist of the exchange couplings between nearest-neighbored MBT-to-MBT, MnTe-to-MnTe, and MBT-to-MnTe layers, respectively. Accordingly, the total energy $E_N$ of the system can be expressed as a function of the applied magnetic field $\boldsymbol{H}$ [36-38]

$$E_N(\boldsymbol{m}_i^{MBT}, \boldsymbol{m}_i^{MnTe}, \boldsymbol{H}) = \sum_{i=1}^{N-1} \left[ J_i^{MBT} \boldsymbol{m}_i^{MBT} \cdot \boldsymbol{m}_{i+1}^{MBT} + J_i^{MnTe} \boldsymbol{m}_i^{MnTe} \cdot \boldsymbol{m}_{i+1}^{MnTe} + \tilde{J}_i^{MBT} (\boldsymbol{m}_i^{MBT} \cdot \boldsymbol{m}_{i+1}^{MBT})^2 \right.$$
$$\left. + \tilde{J}_i^{MnTe} (\boldsymbol{m}_i^{MnTe} \cdot \boldsymbol{m}_{i+1}^{MnTe})^2 \right] - \sum_{i=1}^{N} J_i^{ex} \boldsymbol{m}_i^{MBT} \cdot \boldsymbol{m}_i^{MnTe}$$
$$+ \sum_{i=1}^{N} \left[ K_i^{MBT} t_i^{MBT} (\boldsymbol{m}_i^{MBT} \cdot \boldsymbol{z})^2 + K_i^{MnTe} t_i^{MnTe} (\boldsymbol{m}_i^{MnTe} \cdot \boldsymbol{z})^2 \right]$$
$$- \sum_{i=1}^{N} (M_i^{MBT} t_i^{MBT} \boldsymbol{m}_i^{MBT} + M_i^{MnTe} t_i^{MnTe} \boldsymbol{m}_i^{MnTe}) \cdot \boldsymbol{H} \qquad (1)$$

where $J_i^{MBT}$ and $\tilde{J}_i^{MBT}$ ($J_i^{MnTe}$ and $\tilde{J}_i^{MnTe}$) are the bilinear and biquadratic interlayer exchange coupling constants between adjacent MBT (MnTe) layers, $J_i^{ex}$ denotes the MBT/MnTe interfacial magnetic interaction,

$K_i^{MBT}$ ($K_i^{MnTe}$), $t_i^{MBT}$ ($t_i^{MnTe}$), $M_i^{MBT}$ ($M_i^{MnTe}$), $\mathbf{m}_i^{MBT}$ ($\mathbf{m}_i^{MnTe}$) correspond to the uniaxial anisotropy constant, thickness, saturated magnetization, and the unit vector of magnetization direction of the $i$th MBT (MnTe) layer, respectively. The third term accounts for the anisotropy energy, and the fourth term represents the field-induced Zeeman energy from the uncompensated moments. Under the exchange-spring interaction scenario, the intercalated MnTe acts as a soft FM mediating adjacent antiparallel-coupled MBT layers, and the impact of the $i^{th}$ MnTe spacer on the MBT pair ($M_i^{MBT}$ and $M_{i+1}^{MBT}$) can be expressed as the modifications of the effective interlayer exchange coupling ($J_{eff}^{MBT}$) and uniaxial anisotropy ($K_{eff}^{MBT}$)[36]. Therefore, Equation (1) can be further expressed as the simplified formula:

$$E = J_{eff}^{MBT}[\cos(\theta_i - \theta_{i+1}) + \beta \cos^2(\theta_i - \theta_{i+1})] + K_{eff}^{MBT} t_i^{MBT}\left(\sin^2\theta_i + \xi\sin^2\theta_{i+1}\right) -$$
$$H_z t_i^{MBT}(M_i^{MBT}\cos\theta_i + M_{i+1}^{MBT}\cos\theta_{i+1}) \qquad (2)$$

where $\theta_i$ ($\theta_{i+1}$) is the angle deviation between $M_i^{MBT}$ ($M_{i+1}^{MBT}$) and the magnetic easy-axis (*i.e.*, *z*-direction), $\beta$ and $\xi$ are introduced as the scaling factors of the biquadratic coupling and different anisotropy, and $t_i^{MBT}$ is the MBT layer thickness (see Supplementary Information Section 4 for detailed discussion). From this modified linear chain model, it is found that the transition field $H_t$ is affected by the $K_{eff}^{MBT}$ value whereas the saturation field $H_s$ is mainly decided by the $J_{eff}^{MBT}$ magnitude (Supplementary Fig. 4). Consequently, to re-capture the *m*-dependent MR behaviors observed in Fig. 3, our simulation results suggest that on one hand, the insertion of MnTe modulates the magnetic orientation and anisotropy of the MBT layer (*i.e.*, intra-layer MBT coupling), and the calculated ground-state magnetic configuration of the superlattice with canted spin-polarization (right panel of Fig. 5a) is consistent with the PNR data elucidated in Fig. 2. On the other hand, the presence of MnTe-to-MnTe interlayer coupling and induced MnTe/MBT interfacial interaction both help to preserve the effective interlayer MBT coupling strength (*i.e.*, $J_{eff}^{MBT} = f(J_i^{MnTe}, J_i^{MBT}, J_i^{ex})$) with constant $H_s$ and $T_N$ via the exchange spring effect (Supplementary Fig. 5)[39-41]. Here, we should point out that once the MnTe spacer is replaced by a non-magnetic layer (*e.g.*, $Bi_2Te_3$), the FM-associated exchange interactions are absent, leaving the interlayer MBT coupling determined exclusively by the Anderson super-exchange mechanism (*i.e.*, $J_{eff}^{MBT} = J_i^{MBT}$). Considering the negative correlation between the AFM-type $J_i^{MBT}$ and the spacer distance, the resulting $(MBT)(Bi_2Te_3)_n$ counterpart suffers a rapid suppression of $T_N$ and reduction of $H_s$ with increasing $n$[17,19].

Finally, we show the manipulation of the magnetic/spin configuration in the $[(MBT)(MnTe)_m]_N$ system through superlattice structural engineering. In pure MBT thin films, since both the intra-layer FM and inter-layer AFM strength are non-adjustable with the fixed crystalline structure, Fig. 5b shows that the overall MR profile always maintains the GMR-like line-shape with slightly enlarged $H_t$ and $H_s$ fields as the MBT thickness varies from the 2D (4 SL) to quasi-3D (10 SL) region (the sample thicknesses are calibrated by X-ray reflectivity (XRR) in Supplementary Fig. 6 and Table 6). The reduced GMR amplitude is possibly caused by the increased bulk conduction in thicker films[12,16]. On the contrary, with the insertion of the MnTe spacer, the magnetic proximity effect can re-orient the magnetic moments at the MBT/MnTe interface[34,35,42]. Under such circumstance, it is expected that a change in the number of superlattice repeats (*i.e.,* the number of hetero-interfaces) will introduce dimension-dependent features into the related magneto-transport results. Specifically, the emerging double-split butterfly MR curve with intermediate shoulders develop at ±2.5 T in the thin $[(MBT)(MnTe)_{1.75}]_4$ sample, suggesting the existence of a sizable FM moment that is comparable to the AFM order of the MBT/MnTe/MBT/MnTe unit[43]. As the repeat number increases, the added $J_i^{ex}$ at the hetero-interfaces would promote the exchange-spring effect, as emphasized by the enlarged $H_t$ field in Fig. 5c. Given that the adjacent MBT coupling is mediated through the intervening MnTe, the increase of superlattice repeats will cement such long-range interactions and further stabilize the new ground state. Along with the reinforced canted magnetization orientation which modifies the interfacial spin scattering, it is therefore unsurprising that a more prominent AFM-type GMR behavior is obtained in the $N = 10$ sample[8,44].

In conclusion, we have demonstrated that combining ferromagnets with MBT facilitates the construction of intrinsic MTIs with configurable electronic structures and magnetic properties. In the FM-intercalated MBT systems, the versatile interfacial and interlayer magnetic exchange interactions offer an effective method to shape the overall spin texture, and the AFM-to-FM transition can be well-controlled by structural engineering. Unlike the $(MBT)(Bi_2Te_3)_n$ compounds, the insertion of the MnTe spacer triggers the exchange spring effect which plays an indispensable role in the preservation of the global AFM coupling between adjacent MBT layers throughout the superlattice structures, and this advantage endows the system to be robust against external perturbation, which is favorable for device stability considerations. More importantly, the MR responses can also be tailored via superlattice periodicity optimization, thus providing another degree of freedom in the design of practical spintronic memory/sensor prototypes over traditional magnetic multilayers. With the further exploration of exotic topological features embedded in the host MBT matrix, the $[(MBT)(MnTe)_m]_N$ may serve as an advanced platform for realizing axion insulator states and MTI-based exotic quantum phenomena.

**Methods:**

*Sample Fabrication and structural characterization*: The $[(MBT)(MnTe)_m]_N$ superlattice films were grown on the $Al_2O_3(0001)$ substrate by MBE under a vacuum level of $1 \times 10^{-8}$ Pa. Prior to the sample growth, the $Al_2O_3$ substrate was pre-annealed at 570 °C in order to remove the absorbed contamination. The growth temperature for $Bi_2Te_3$ and MnTe were kept at 200 °C and 370 °C, respectively. After post-annealing at 390 °C for 3 minutes, a period of the $MBT(MnTe)_m$ film was fabricated. During the MBE growth, high-purity Mn and Bi atoms were evaporated from standard Knudsen cells, while Te was evaporated by a thermal cracker cell. Meanwhile, we used a beam flux monitor to measure the flux ratio and RHEED to monitor the real-time growth condition. Besides, X-ray diffraction and reflectivity were performed to measure the crystal structure and calibrate the film thickness, and Atomic-resolution structural characterization was performed by aberration-corrected Hitachi HF5000 STEM/TEM under the high-angle annular dark field (HAADF) STEM mode, operated at 200 kV. Cross-section TEM specimen was prepared by FEI Scios FIB.

***Transport Measurements***: The as-grown superlattices were etched into a six-probe Hall bar geometry with a dimension of 2 mm × 1 mm. The electrodes were made by welding small pieces of indium onto the contact areas. The magneto-transport measurements of the fabricated devices were performed with the $He^4$ refrigerators (Oxford Teslatron PT system) where several experimental variables such as temperature, magnetic field, and lock-in frequency were adjusted during the measurements. Multiple lock-in amplifiers and Keithley source meters (with an excitation AC current $I = 1$ μA) were connected to the samples so as to enable the precise four-point lock-in experiments.

***Neutron Reflectometry:*** Unpolarized neutron reflectometry measurements and polarized neutron reflectometry measurements were performed using the PBR instrument at the NIST Center for neutron research. Incident and scattered neutrons were polarized parallel or antiparallel to the direction of the applied magnetic field. Due to the 3 T applied field, no net perpendicular magnetization is expected in the plane of the film, so that the spin-flip scattering cross-sections may be neglected. We consequently measured the non-spin-flip scattering cross-sections as a function of $Q$, the momentum transfer vector along the film normal direction. The data was reduced using the Reductus[45] software package. Further unpolarized neutron reflectometry measurements were performed using the POLREF[46] instrument at the ISIS Neutron and Muon Source[47]. The incident and scattered beams were unpolarized, and the reflectivity was measured at room-temperature, well above any magnetic transition. All reflectivity analysis was performed using the Refl1D software package[48].

**Acknowledgements**

This work is sponsored by the National Key R&D Program of China under the contract number 2017YFB0305400, National Natural Science Foundation of China (Grant No. 61874172 and 11904230), and the Major Project of Shanghai Municipal Science and Technology (Grant No. 2018SHZDZX02). X.F.K acknowledges the support from the Merck POC program and the Shanghai Rising-Star program (Grant No. 21QA1406000). Y.M.Y acknowledges the support from Shanghai Pujiang Program (Grant No. 20PJ1411500). The authors acknowledge the facilities, and the scientific and technical assistance, of the Australian Microscopy & Microanalysis Research Facility at the Centre for Microscopy and Microanalysis, The University of Queensland. Q.Y acknowledges the support from the Shanghai Sailing Program (Grant No. 19YF1433200). We would also like to thank the ISIS neutron facility for the awarding of beamtime RB2000244, DOI: [10.5286/ISIS.E.RB2000244](10.5286/ISIS.E.RB2000244). Certain commercial equipment is identified in this paper to foster understanding. Such identification does not imply recommendation or endorsement by NIST.

**Author contributions**

X. F. Kou, and Q. Yao conceived and supervised the study. P. Chen grew the samples, performed the characterization measurements and conducted the transport measurements. Q. Yao and P. Chen analyzed the transport and characterization data. P. Chen and Y. M. Yang conducted the macro-spin simulations. Q. Sun and J. Zou performed microscopy characterization. A. Grutter, P. Quarterman, P. P. Balakrishnan, C. Kinane, A. Caruana, and S. Langridge, performed the neutron reflectometry measurements. P. Chen, Q. Yao, Y. M. Yang and X. F. Kou wrote the manuscript. All authors discussed the results and commented on the manuscript.

**Competing financial interests**

The authors declare no competing financial interests.

# Figures

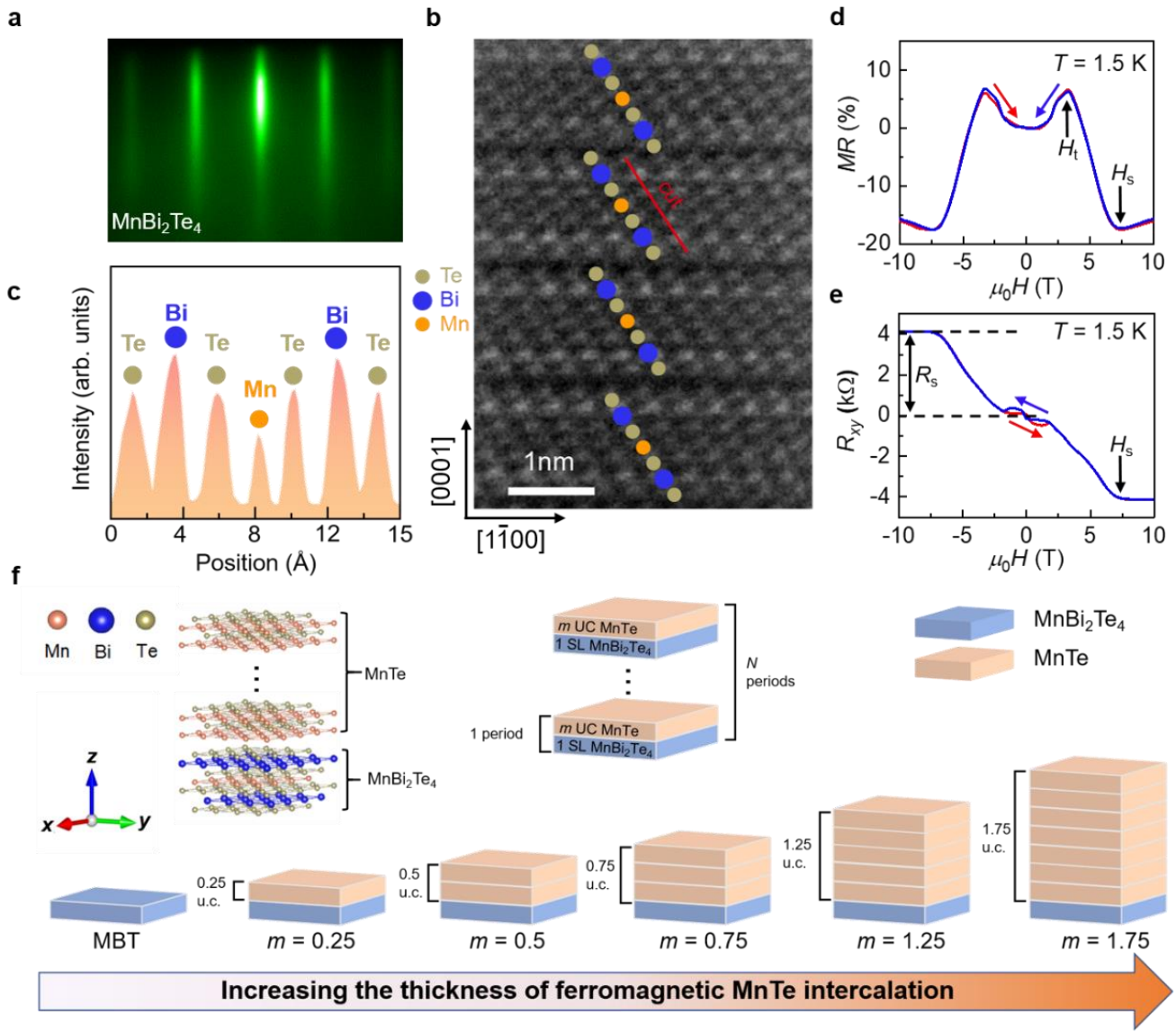

**Figure 1. Structural and electrical characterizations of the MBE-grown MBT thin film**. **a**, *In-situ* RHEED with a sharp streaky pattern reveals the epitaxial growth mode. **b**, Cross-sectional HR-STEM image of the MBT film grown on the $Al_2O_3$ (0001) substrate. **c**, EELS intensity mapping along the cut marked in **b**. **d-e**, Magnetic field-dependent MR and $R_{xy}$ data of the 5 SL MBT sample. Both the AFM-type GMR line-shape and spin-flop-induced hysteresis loop at 1.5 K are consistent with bulk MBT data. **f**, Schematic diagram of the proposed FM-intercalated $[(MBT)(MnTe)_m]_N$ structure. By varying the MnTe thickness ($m$) and superlattice repeats ($N$), the magnetic exchange interactions can be regulated.

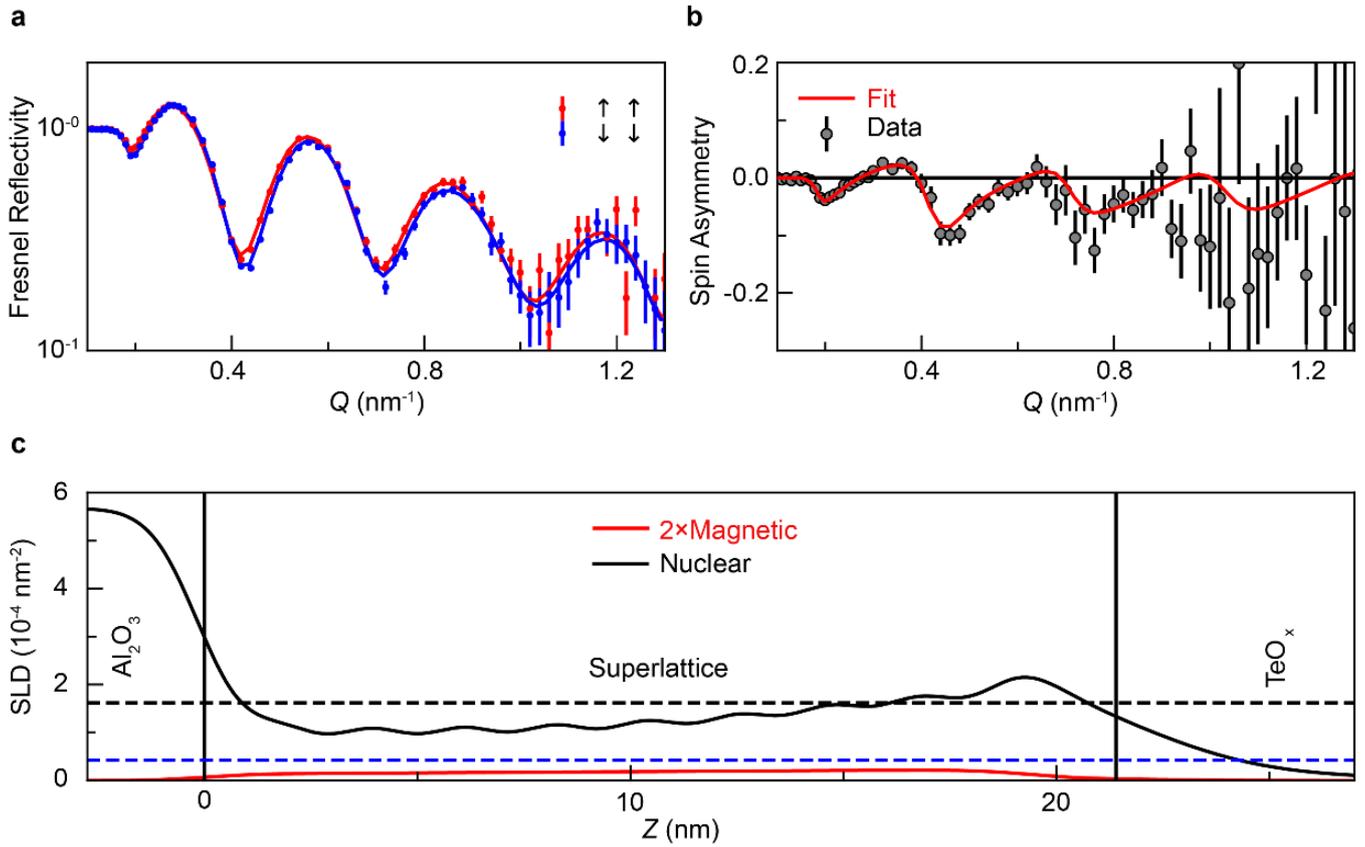

**Figure 2. Polarized neutron reflectometry measurements on the MBT-grown [(MBT)(MnTe)$_{1.75}$]$_{10}$ superlattice. a-b**, Polarized neutron reflectivity and spin asymmetry alongside best fit for the Te-capped [(MBT)(MnTe)$_{1.75}$]$_{10}$ sample. **c**, Scattering length density (SLD) profile used to generate the fit. Error bars represent ±1 standard deviation. Measurements are performed at $T$ = 6 K under a 3 T in-plane applied magnetic field. The black dashed line represents the expected bulk nuclear SLD of the MBT component while the blue dashed line corresponds to the ideal bulk nuclear SLD of MnTe. An in-plane magnetization of around 20 emu/cc (averaged over the whole superlattice) is obtained from the PNR data.

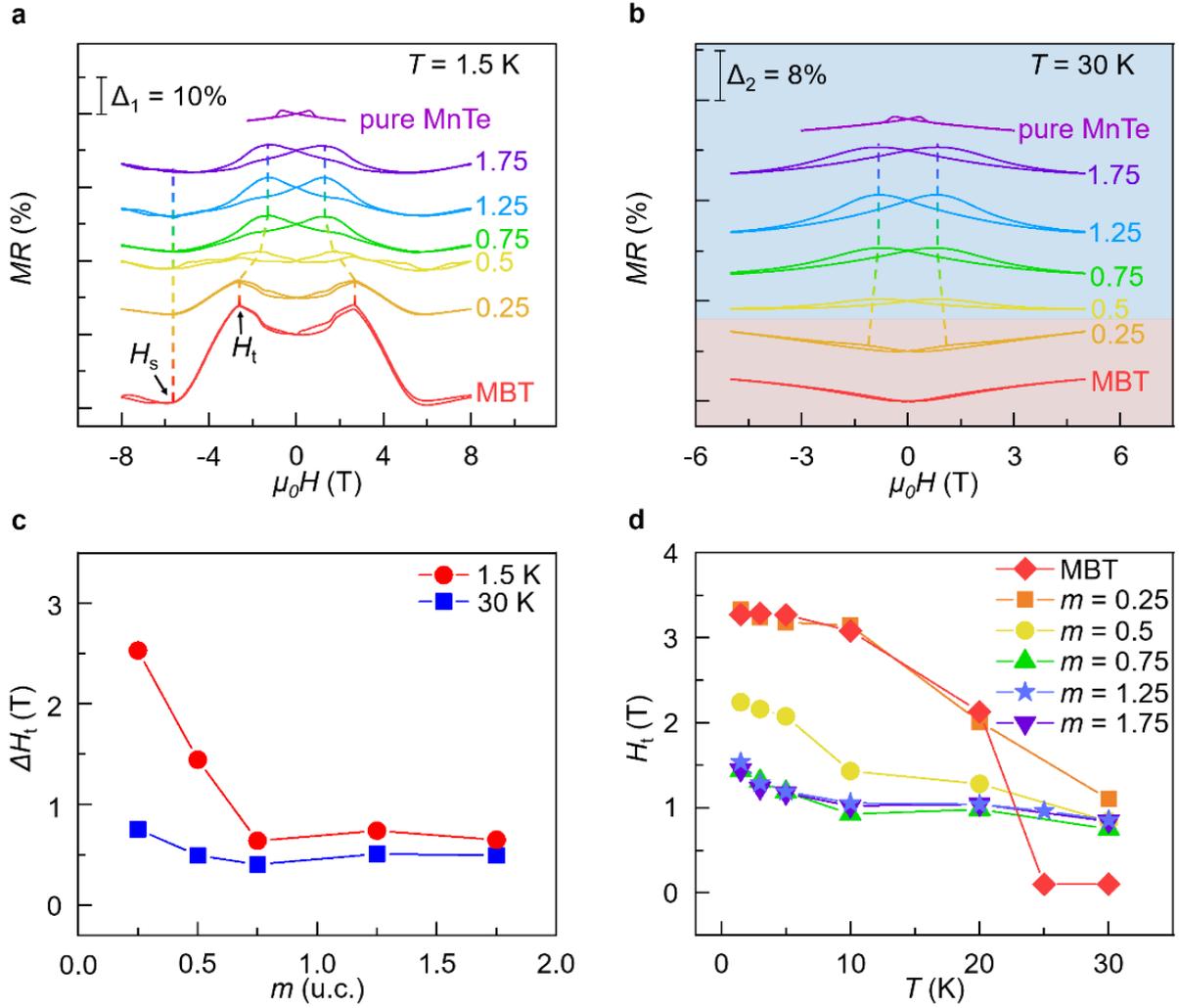

**Figure 3. Manipulation of the exchange spring effect in [(MBT)(MnTe)$_m$]$_5$ samples through MnTe intercalation.** MnTe spacer thickness-dependent MR curves at **a**, $T$ = 1.5 K and **b**, $T$ = 30 K, respectively. The value of $m$ varies from 0 to 1.75. Data are shifted vertically by $\Delta_1$ and $\Delta_2$ for convenient comparison. **c**, The evolutions of $\Delta H_t$ with respect to the MnTe thickness. The negative correlation between $\Delta H_t$ and $m$ at 1.5 K (red circles) manifests the exchange spring effect whereas the constant $\Delta H_t$ values at 30 K (blue squares) confirm the absence of the AFM order in the system above the Néel temperature. **d**, The temperature-dependent $H_t$ traces of the [(MBT)(MnTe)$_m$]$_5$ samples exhibit a clear MBT-to-MnTe transition feature at low temperatures.

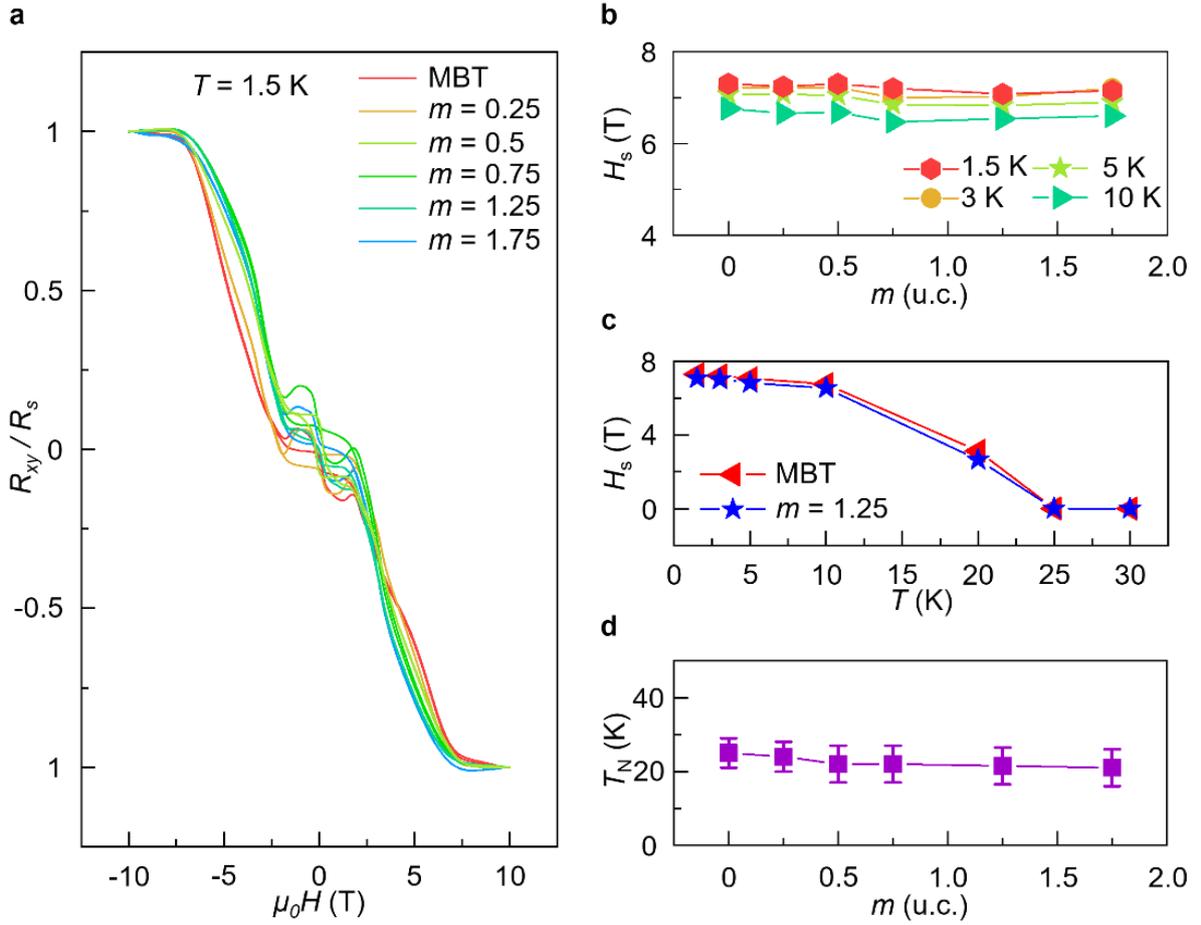

**Figure 4. Mediation of interlayer couplings in the [(MBT)(MnTe)$_m$]$_5$ system. a**, $m$-dependent anomalous Hall curves at $T$ = 1.5 K. All samples show the same spin-flop behavior regardless of the MnTe thickness. **b**, The extracted transition field $H_s$ is insensitive to the $m$ at different temperatures. **c**, Temperature-dependent $H_s$ curve of the [(MBT)(MnTe)$_{1.25}$]$_5$ superlattice as compared with the pure 5 SL MBT control sample ($m$ = 0). **d**, The Néel temperature $T_N$ remains almost unchanged in all [(MBT)(MnTe)$_m$]$_5$ samples, indicating the stabilization effect by the MnTe intercalation.

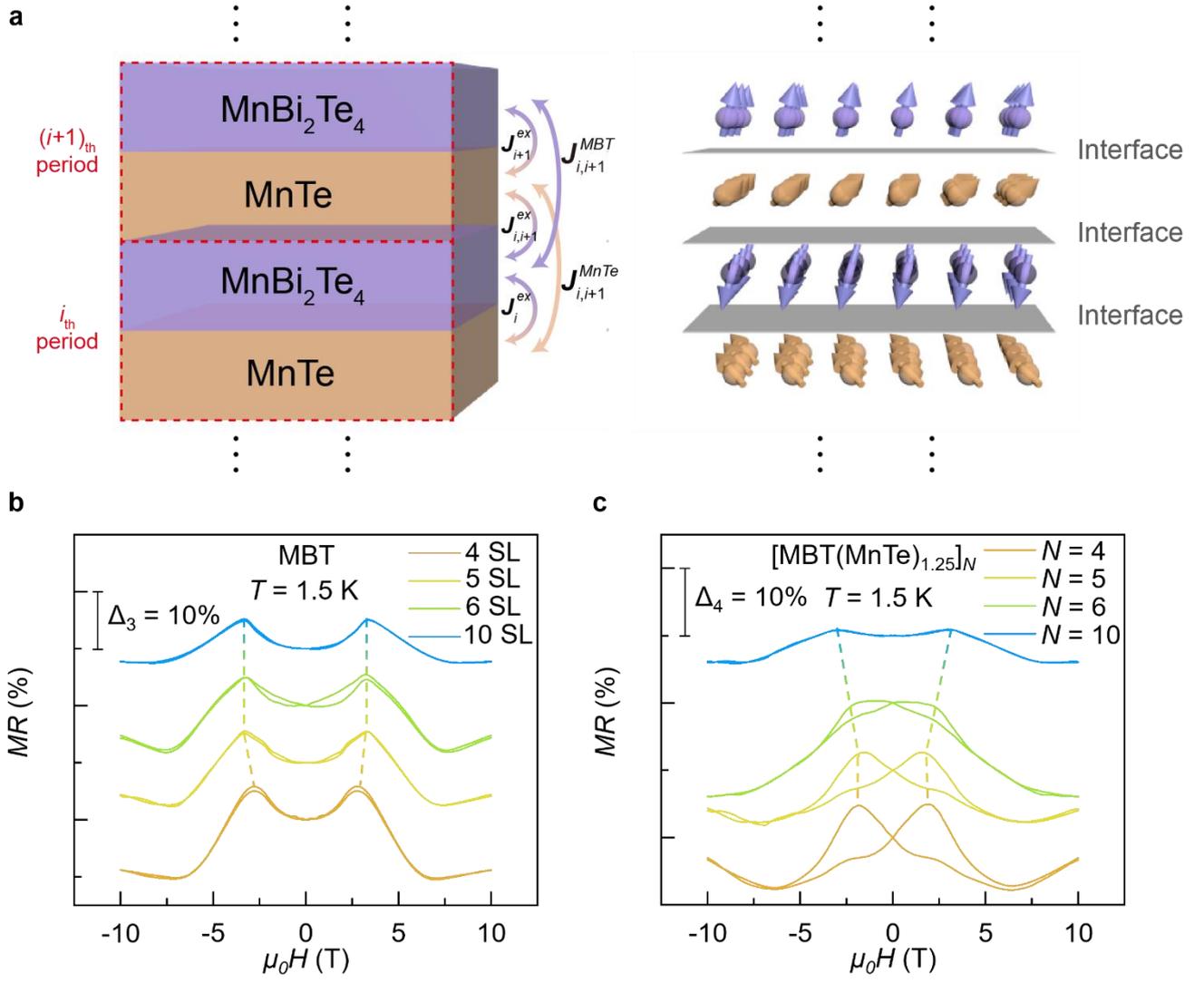

**Figure 5. Tunable magneto-resistance responses through superlattice engineering. a**, Illustrations of the interfacial and interlayer exchange interactions (left panel) and the resulting canted spin configuration (right panel) for the [(MBT)(MnTe)$_m$]$_N$ superlattices based on the modified linear chain model simulation. The in-plane spin rotation is induced by $J_i^{ex}$. Evolutions of MR responses in **b**, thickness-dependent pure MBT thin films and **c**, [(MBT)(MnTe)$_{1.25}$]$_N$ samples. With the number of superlattice repeats $N$ increases from 4 to 10, the reinforced canted magnetization and long-range interactions drive the overall MR signal from the FM-dominated double-split butterfly shape towards the AFM-type GMR contour.